\documentstyle[11pt]{article}
\begin{document}
\thispagestyle{empty}
\begin{flushright}
MPI-PhT 96-18 \\
UNIGRAZ-UTP-13-03-96\\
hep-lat/9603010
\end{flushright}
\vskip10mm
\begin{center}
{\huge{Lattice Fermions Coupled to Interpolated \vskip3mm
Gauge Fields - Results in 2 Dimensions}${}^*$}
\vskip10mm
\centerline{ {\bf
Christof Gattringer${}^{**}$ }}
\vskip 5mm
\centerline{Max-Planck-Institut f\"{u}r
Physik (Werner-Heisenberg-Institut)}
\centerline{F\"ohringer Ring 6, 80805 Munich, Germany}
\vskip 2mm
\centerline{and}
\vskip 2mm
\centerline{Institut f\"{u}r Theoretische Physik der Universit\"at Graz} 
\centerline{Universit\"atsplatz 5, 8010 Graz, Austria}
\vskip30mm
\end{center}
\begin{abstract}
\noindent
We discuss a new approach for putting gauge theories on the lattice. 
The gauge fields are defined on the lattice only, but are interpolated to the 
interior of the lattice cells, where they couple to continuum fermions.
The purpose of this approach is to avoid doublers and keep the chiral
symmetry of the action intact.
In two dimensional models (Schwinger model and chiral Schwinger model) 
many results for this hybrid approach can be obtained analytically. For the 
vectorlike Schwinger model we concentrate on proving the existence of 
a critical point where the continuum limit can be constructed, on analyzing 
the chiral properties and on proving the Osterwalder-Schrader positivity 
which allows the reconstruction of the physical Hilbert space. We conclude 
with outlining how the results can be generalized to the chiral Schwinger 
model.
\end{abstract}
\vskip20mm
\bigskip \nopagebreak \begin{flushleft} \rule{2 in}{0.03cm}
\\ {\footnotesize 
\noindent 
${}^*$ Talk given at `Hadron Structure 96', February, 11-16, 1996, Stara Lesna,
Slovakia \\
${}^{**}$ e-mail: chg@mppmu.mpg.de}
\end{flushleft}
\newpage
%
%
%
\section{Introduction}
It has been known since the first days of lattice field theory, 
that maintaining
the chiral properties of lattice fermions is a deep problem
(see e.g. Shamir 1995 for a review).
Either one encounters unwanted fermionic degrees of freedom (doublers)
or one destroys the chiral symmetry of the action when removing the 
doublers by introducing a Wilson term. A possible way out 
of the dilemma (`hybrid approach')
is to put the gauge fields on the lattice, interpolate them to the interior 
of the lattice cells and couple them to continuum fermions 
(Flume \& Wyler 1982; Stuller; G\"ockeler \& Schierholz 1992, 1993; 
't Hooft 1995; Hsu 1995; Montvay 1995;
Hernandez \& Sundrum 1995a,b, 1996; Gattringer 1995).
't Hooft (1995) argued that for U(1) gauge fields with vectorlike coupling 
to the fermions the fermion determinant with interpolated lattice fields 
can be given a meaning, that all chiral symmetries are kept intact and that 
no doubling occurs. This is an important result, but unfortunately 
it is not possible to analyze details analytically. In particular one is 
interested in identifying critical points where the continuum limit can be 
reconstructed and in proving Osterwalder-Schrader (OS) positivity which 
guarantees that a physical Hilbert space can be defined.
Those more fundamental questions can be addressed in two 
dimensional models. 
        
In this contribution we discuss how the hybrid approach works out for
the Schwinger model and for the chiral Schwinger model. For the vectorlike
Schwinger model we prove the existence of a critical point with diverging 
correlation length, discuss the chiral properties and prove OS-positivity.

\section{Interpolation and effective action}
The lattice under consideration is
$\mbox{Z\hspace{-1.35mm}Z}^2$, i.e. the lattice spacing is set to one.
Functions defined in the continuum can be identified by their 
arguments $x,y \in \mbox{I\hspace{-0.7mm}R}^2$, 
while lattice quantities have arguments
$n,m,k,r,s \in \mbox{Z\hspace{-1.35mm}Z}^2$.
For the gauge field action we consider the non-compact formulation
\begin{equation}
S_{G} \; := \; \frac{1}{4} \sum_{{n} \in Z\mbox{\hspace{-1.35mm}}Z^2}
\; F_{ \mu \nu}({n}) F_{ \mu \nu}({n}) \; = \; 
\frac{1}{2} \sum_{{n} \in Z\mbox{\hspace{-1.35mm}}Z^2}
\; F_{12}({n})^2 \; \;  ,
\end{equation}
with
\begin{equation}
F_{\mu \nu}({n}) \;  := \;  \Big( A_\nu({n} + \hat{e}_\mu) -
A_\nu({n}) - A_\mu({n} + \hat{e}_\nu) + 
A_\mu({n}) \Big) \; .
\end{equation}
$A_\mu({n})$ may assume values in $(-\infty, + \infty)$ for all 
${n} \in \mbox{Z\hspace{-1.35mm}Z}^2$ 
and $\mu = 1,2$. $\hat{e}_\mu$ is the unit
vector in $\mu$-direction. The gauge field action (1) is
invariant under the lattice 
gauge transformation
\begin{equation}
A_\mu({n}) \; \longrightarrow \; A_\mu({n}) \; \; \; + 
\; \Lambda({n} + \hat{e}_\mu) - 
\Lambda({n})  \; ,
\end{equation}
where $\Lambda({n})$ is some scalar lattice field.
Having fixed the discretization of the gauge field action and
thus the form of the gauge transformation on the lattice (3),
we can start thinking about an interpolation to the interior of the
lattice cells. A basic requirement for the interpolation is 
that the continuum fermion determinant in the external, interpolated
gauge field has to be invariant under the lattice gauge transformations (3). 
We can use the gauge invariance of the continuum determinant under
continuum gauge transformations to fulfill this requirement. We use an 
interpolation prescription for the lattice fields $A_\mu(n)$,
$\Lambda(n)$ which transforms the lattice gauge transformation (3) into
a continuum transformation 
$A_\mu(x) \rightarrow A_\mu(x) + \partial_\mu \Lambda(x)$ 
for the interpolated fields $A_\mu(x), \Lambda(x)$. A possible choice, 
respecting this condition is given by (see Gattringer (1995) for details)
\begin{equation}
A_1({x}) \; := \; A_1({n}) \;  [1-t_2]  \;  + \;  
A_1({n}+\hat{e}_2) t_2 \; \; \; \; , \; \; \; \; 
A_2({x}) \; := \; A_2({n}) \;  [1-t_1]  \;  + \;  
A_2({n}+\hat{e}_1) t_1 \; \; ,
\end{equation}
\begin{equation}
\Lambda({x}) \; := \; \Big( \Lambda({n}) [1-t_1] + 
\Lambda({n}+\hat{e}_1) t_1 \Big) [1-t_2] \; + \; 
\Big( \Lambda({n}+\hat{e}_2) [1-t_1] + 
\Lambda({n}+\hat{e}_1+\hat{e}_2) t_1 \Big) t_2 \; \; ,
\end{equation}
for ${x} = {n} + {t} $ and $t_1,t_2 \in (0,1]$. It is 
easy to check that the interpolation (4), (5) obeys the requirement of 
transforming a lattice gauge transformation into a continuous one. The 
interpolation (4) was already obtained earlier
(Flume \& Wyler 1982, G\"ockeler et al. 1993);
but the motivation was different, namely the 
definition of a winding number for lattice gauge fields (L\"uscher 1982).
It has to be remarked, that the interpolated 
gauge fields $A_\mu(x)$ are not differentiable on the links 
and in $\mu$-direction even not continuous. In particular the 1-component 
$A_1$ is discontinuous with finite step height
on the links parallel to 2-direction and vice versa.
Thus the derivatives of the $A_\mu$ have to be computed in the sense of 
distributions giving rise to $\delta$-distributions times the step height
at the lines of discontinuity.
However, things work out simpler if one considers gauge invariant 
operators only, i.e. functions of $F_{12}({n})$. 
Since $A_1$ is continuous in 2-direction and vice versa, the field strength 
$F_{12}$ picks up no contributions involving $\delta$-distributions
\begin{equation}
F_{12}({x}) 
\; = \; \Big( A_2({n} + \hat{e}_1) -
A_2({n}) - A_1({n} + \hat{e}_2) + 
A_1 ({n}) \Big) \; = \; F_{12}({n}) \; ,
\end{equation}
for ${x} = {n} + t $ and $t_1,t_2 \in (0,1]$.
The interpolated field strength is constant inside the lattice cells,
but discontinuous along the links.
For later use we quote the continuum Fourier transform 
$\widetilde{F_{12}}({p})$
of the interpolated field strength given by (6)
\begin{equation}
\widetilde{F_{12}}({p}) \; 
:= \; \int_{-\infty}^{\infty} d^2 x \; F_{12} ({x})
e^{-i {p} \cdot {x}} 
\; = \; - \widehat{F_{12}}({p}) \; \frac{1-e^{-ip_1}}{ p_1} \; 
\frac{1-e^{-ip_2}}{ p_2} \; ,
\end{equation}
where we introduced the lattice Fourier 
transform $\widehat{F_{12}}({p})$ as
\begin{equation}
\widehat{F_{12}}(p) \; := \;
\sum_{{n} \in Z\hspace{-1.35mm}Z^2} \; F_{12}({n}) \;
e^{-i{p} \cdot {n}} \; .
\end{equation}
$\widehat{F_{12}}({p})$ is periodic in ${p}$ with respect to the 
Brillouin zones.

The next step in the hybrid approach is to give a meaning to the fermion
determinant in the continuum and plug in the interpolated gauge field. 
For the Schwinger model the regularized continuum determinant in an
external field can be computed easily (Schwinger 1962)
\begin{equation}
\mbox{det}_{reg}[ 1 - K(A) ] \; = \; \exp \left( -\frac{e^2}{2\pi} 
\int_{-\infty}^{\infty} \frac{ d^2 p}{(2\pi)^2} \;   
\widetilde{F_{12}}(- {p}) \;
\frac{1}{{p}^{2}} \; \widetilde{F_{12}}({p}) \right) \; .
\end{equation}
It has to be remarked that this expression makes sense only if
the gauge field satisfies some mild regularity and falloff conditions.
In particular $\widetilde{F_{12}}(p)$ has to vanish for zero momentum.
This corresponds to zero topological charge. For the lattice fields
this implies (compare (7)) $0 =  
\widehat{F_{12}}({0}) =  
\sum_{{n}} [ A_2({n} + \hat{e}_1) -
A_2({n}) - A_1({n} + \hat{e}_2) + 
A_1({n}) ]$. The last equation is always fulfilled if the 
lattice gauge fields $A_\mu({n})$ are absolutely summable over 
Z\hspace{-1.35mm}Z$^2$. This is the announced falloff condition expressed 
in terms of the lattice gauge fields. It can e.g. be imposed by
restricting the $A_\mu({n})$ to a finite rectangle in
Z\hspace{-1.35mm}Z$^2$. 
Inserting (7) in (9) one obtains the contribution of the fermion 
determinant to the effective lattice gauge field action
\begin{equation}
S_F \; := \; \frac{e^2}{2\pi} 
\int_{-\pi}^{\pi}
\frac{d^2p}{(2\pi)^2} \; \widehat{F_{12}}(-{p}) \sigma(p)
\widehat{F_{12}}({p}) \; ,
\end{equation}
where $\sigma(p)$ is given by
\begin{equation}
\sigma(p) \; := \;
\sum_{{k} \in Z\hspace{-1.35mm}Z^2}
\frac{1}{({p} + 2\pi {k}\,)^2} \;
\frac{2-2\cos( p_1)}{(p_1 +  2 \pi k_1)^2} \;
\frac{2-2\cos( p_2)}{(p_2 +  2 \pi k_2)^2} \; .
\end{equation}

Fourier transforming the gauge field action (1) 
and adding it to (10) gives the effective lattice action  
\begin{equation}
S_{EFF} := 
\frac{1}{2}
\int_{-\pi}^{\pi} \frac{d^2p}{(2\pi)^2} \; \widehat{F_{12}}(-{p}\,) 
\widehat{F_{12}}({p}\,) \left[1+
\frac{e^2}{\pi}\sigma(p)\right] \; .
\end{equation}
The effective action for non-compact gauge fields is a quadratic form 
in the field strength $F_{12}$.
As in the continuum, when restricting to gauge
invariant observables, the model can be quantized by defining a path integral 
for $F_{12}$ (Gattringer \& Seiler 1994) and no extra gauge fixing term has to 
be taken into account. This gives rise to a
Gaussian measure $d\mu_C[F_{12}]$ with lattice-covariance
\begin{equation}
\widehat{C}({p} ) \; := \; 
\left[ \; 1 \; + \; \frac{e^2}{\pi} \sigma(p) \right]^{-1} \; .
\end{equation}
The Gaussian measure $d\mu_C[F_{12}]$ with covariance
$C$ is the starting point for quantizing the effective lattice gauge theory.

\section{Correlation length and continuum limit}
In order to compute the correlation length of the effective lattice
gauge theory, we evaluate the two point function of the field strength.
It will be shown that for small $e$ the two point function falls
off exponentially for large time-like separation, thus defining a 
correlation length. The two point function of $F_{12}({n})$ is simply
given by the inverse Fourier transform of the covariance (13)
\[
\langle F_{12}(0,t)  F_{12}(0,0) \rangle = 
\int_{-\pi}^{\pi} \frac{d^2p}{(2\pi)^2} e^{-ip_2t} 
\widehat{C}({p})  
= \int_{-\pi}^{\pi} \frac{d^2p}{(2\pi)^2}  e^{-ip_2t}  
\left[ 1 - \frac{e^2/\pi}{\sigma(p)^{-1} +
\frac{e^2}{\pi} } \right] 
\]
\begin{equation}
= \; \delta_{0,0} \delta_{0,t} \; - \; \frac{e^2}{\pi} 
\int_{-\pi}^{\pi} \frac{d^2p}{(2\pi)^2} \; e^{-ip_2t}
\frac{1}{\sigma(p)^{-1} +
\frac{e^2}{\pi} } \; .
\end{equation}
In the second step the contact term which is also known from 
the continuum was split of.
In order to extract the exponential falloff of the two point function,
the singularities of $f(p) \; := \; \big[ \sigma(p)^{-1} +
e^2/\pi \big]^{-1}$
in the complex $p_2$-plane have to be determined. Since $\sigma(p)$ contains
the cosine factors, one would have to solve transcendental equations, which
cannot be done in closed form. Thus one is reduced to a perturbative analysis 
for small gauge coupling $e$. As can be seen from (11),
$\sigma(p)^{-1}$ behaves as $p_1^2 + p_2^2$ for small $p$. Thus for small 
$e$ and $p$ the integrand $f(p)$ behaves as 
$\big[p^2 + e^2/\pi \big]^{-1}$.
This behaviour suggests for the correlation length $\xi$ 
(see Gattringer (1995) for a careful analysis of $f(p)$)
\begin{equation}
\xi \; = \; \frac{\sqrt{\pi}}{e} \; .
\end{equation}
Of course this perturbative analysis does not proof (15).
However for this simple model one can do better, 
in particular the continuum limit can be controlled analytically. So we 
assume for the moment that $\xi$ is given by (15). This implies that the 
continuum limit can be obtained by  $e \rightarrow 0$, where 
the correlation length becomes infinite. 
We now define our length scale $L_0$ to be proportional to the correlation
length, i.e~$L_0 := \lambda \xi$. A physical distance $|x|$ is 
measured in units of $L_0$ giving rise to $|x| = t/L_0$. 
The continuum gauge coupling $e_c$ which has the dimension 
of a mass is defined as $e_c = eL_0$. Thus we obtain 
for the ratio $t/\xi$ 
\begin{equation}
\frac{t}{\xi} \; = \; t \frac{e}{\sqrt{\pi}} \; = \; 
\mbox{const} \; = \; |{x}| \frac{e_c}{\sqrt{\pi}} \; .
\end{equation}
Thus the continuum limit is the joint limit of letting $e \rightarrow 0$ 
and keeping the product $et$ fixed, where $t$ is the distance of the
operators on the lattice (compare (14)). In Gattringer (1995) it was shown 
that the overal factor $e^2$ in front of the integral in the last term 
of (14) has to be removed with a wave function renormalization constant 
$Z_{12}(e) = e^2_c/e^2$ when performing the limit $e \rightarrow 0$. 
Using this renormalization procedure and performing the continuum limit
one ends up with
\begin{equation}
\lim_{e \rightarrow 0 \atop et = const} \; Z_{12}(e) 
\langle F_{12} (0,t) \; F_{12} (0,0) \rangle \; = 
\langle F_{12}(x) \; F_{12}(0) \rangle_{cont} \;,
\end{equation}
where the continuum expression 
$\langle F_{12}(x) F_{12}(0) \rangle_{cont}$ is given by (see e.g.
Seiler \& Gattringer 1994)
\begin{equation}
\langle F_{12}(x) F_{12}(0) \rangle_{cont} \; = \; 
\delta^{(2)}({x}) \; - \; \frac{e_c^2}{\pi} \frac{1}{2 \pi}
\mbox{K}_0 \Big(\frac{e_c}{\sqrt{\pi}} |{x}| \Big) \; .
\end{equation}
K$_0$ denotes the modified Bessel function. The proof of (17), (18) 
consists of
two steps. The contact term of (18) is recovered from the (lattice) contact
term in (14) since (note the wave function renormalization) 
$e_c^2/e^2 \delta_{0,t} \delta_{0,0} \rightarrow \delta^{(2)}(x)$ 
when taking the 
limit $e \rightarrow 0$ in the sense of (16).
The convergence of the second term in (14) can 
be proven by using the Riemann Lebesgue lemma to show that
\begin{equation}
\int_\pi^\pi \frac{d^2p}{(2\pi)^2} \frac{e^{-ip_2t}}{\sigma(p)^{-1} + 
\frac{e^2}{\pi}} \; - \; \int_\pi^\pi \frac{d^2p}{(2\pi)^2} 
\frac{e^{-ip_2t}}{p^2 + \frac{e^2}{\pi}} \; ,
\end{equation}
vanishes when $e \rightarrow 0$ with the restriction (16) (see Gattringer
1995 for an explicit proof). The second term in (19) gives for 
$e \rightarrow 0$
the modified Bessel function showing up in the right hand side
of (18). Thus the continuum limit for the two point function
of the field strength is under control.

\section{Chiral properties}
For the continuum Schwinger model with N-flavors the simple N=1 condensate 
$\langle \overline{\psi} \psi \rangle$ is replaced by a higher monomial
of the fermions. In particular for the N-flavor model obtains
(Belvedere et al. 1979, Gattringer \& Seiler 1994) 
\begin{equation}
\lim_{|{x} - {y}| \rightarrow \infty}\!\Big\langle
\prod_{a=1}^j \overline{\psi}^{(a)}(x) P_+ \psi^{(a)}(x) \; 
\overline{\psi}^{(a)}(y) P_- \psi^{(a)}(y) \Big\rangle_{cont}\!=
\left\{ \begin{array}{cc}
\Big(-\frac{e_c^2N}{16\pi^3}e^{2\gamma} \Big)^N & \mbox{for} 
\; j = N \\ 0 & \mbox{for} \; j < N
\end{array} \right. 
\end{equation} 
where $a$ is the flavor index, $P_\pm$ denote the projectors on left and 
right handed chirality, $\gamma$ is Euler's constant and finally $e_c$ denotes
the gauge coupling of the continuum model.
The chiral condensate of the N-flavor model 
involves chiral densities of all flavors, and thus is sensitive to the number
of flavors. One can learn about the doubling problem and the chiral
properties of the lattice
model by comparing it's condensate to the condensate of the N-flavor 
continuum model. On the lattice we will evaluate
\begin{equation}
\langle \overline{\psi}({n}) P_+ \psi({n}) \;
\overline{\psi}({m}) P_- \psi({m}) \rangle \; = \;
- \int d \mu_C [ F_{12}] \; G_{12}({n},{m};A_\mu)
G_{21}({m},{n};A_\mu) \; .
\end{equation}
$G$ is the continuum fermion propagator in an external field given by
(Schwinger 1962)
\begin{equation}
G({n},{m};A_\mu) \; = \;  \frac{1}{2\pi} \frac{\gamma_\mu 
(n_\mu - m_\mu)}{({n} - {m})^2} \;
e^{i[\Phi({n}) - \Phi({m})]}  \; ,
\end{equation}
with
\begin{equation}
\Phi({n}) \; = \; -\int d^2x  D({n}-{x}) 
\Big( \partial_\mu A_\mu(x) + i \gamma_5
\varepsilon_{\mu \nu} \partial_\mu A_\nu({x}) \Big) \; =: \;
\theta({n}) + i \gamma_5 \chi({n}) \; , 
\end{equation}
and $D$ denotes the Green's function of $-\triangle$.
In the last step we introduced the longitudinal part
$\theta = \triangle^{-1} \partial_\mu A_\mu$ which cancels in
gauge invariant expectation values like (21), and the gauge invariant 
part
\begin{equation}
\chi({n}) \; := \; \triangle^{-1} F_{12}({n}) = 
\int_{-\pi}^{\pi} \frac{d^2p}{(2\pi)^2} \; \widehat{F_{12}}({p}\,)
e^{i{p}\cdot{n}} 
\sum_{{k} \in Z\hspace{-1.35mm}Z^2} \; 
\frac{1}{({p}+2\pi{k})^2} \;
\frac{1-e^{-ip_1}}{ p_1 + 2\pi k_1} \; 
\frac{1-e^{-ip_2}}{ p_2 + 2\pi k_2} \; ,
\end{equation}
where the Fourier transform (7) of the interpolated $F_{12}$ was inserted.
The Gaussian integral in (21) can be solved, giving rise to
(for simplicity we choose the space-time arguments to be ${n} = (0,t),
{m} = (0,0)$)
\begin{equation} 
\langle \overline{\psi}(0,t) P_+ \psi(0,t) \;
\overline{\psi}(0,0) P_-\psi(0,0) \rangle \; = \;
- \; \frac{1}{(2\pi)^2} \; \frac{1}{t^2}\; e^{2E(t,e)} \; ,
\end{equation}
where we defined
\begin{equation}
E(t,e) \; = \; e^2
\int_{-\pi}^{\pi} \frac{d^2p}{(2\pi)^2} \; [2-2\cos(p_2t)] \;
\frac{\rho(p)}{1 + \frac{e^2}{\pi} \sigma(p)} \; .
\end{equation}
$\rho(p)$ is given by 
\begin{equation}
\rho(p) := \sum_{{r},{s} \in Z\hspace{-1.35mm}Z^2} \!
\frac{1}{({p}\!+\!2\pi{r})^2} 
\frac{1}{({p}\!+\!2\pi{s})^2} 
\frac{2-2\cos(p_1)}{(p_1\!+\!2\pi r_1)(p_1\!+\!2\pi s_1)}  
\frac{2-2\cos(p_2)}{(p_2\!+\!2\pi r_2)(p_2\!+\!2\pi s_2)} \; .
\end{equation}
The condensate is now being formed by $E(t,e)$ which contains a 
term proportional to $\ln(t)$ which cancels the $1/t^2$ factor in (25),
plus a term which approaches a finite constant for $t \rightarrow \infty$.
In Gattringer (1995) it is shown that the exponent $E(t,e)$ behaves as
$\ln(t) + R(t,e)$,
where $\lim_{t \rightarrow \infty} R(t,e) = r(e)$, and $r(e)$ is some 
finite constant. Thus we end up with 
\begin{equation}
\lim_{t \rightarrow \infty} 
\langle \overline{\psi}(0,t) P_+ \psi(0,t) \;
\overline{\psi}(0,0) P_-\psi(0,0) \rangle\; = \;
- \; \frac{1}{(2\pi)^2} \; e^{2 r(e)} \; \neq \; 0 \; .
\end{equation}
The result (28) is exact for all $e > 0$ and contains no expansion in $e$.
Thus the one flavor condensate already forms for $e > 0$ i.e. at finite 
correlation length.

The nonvanishing condensate for finite $e$ is an important result.
If there were doublers in the model under consideration, the chiral
condensate would involve higher powers of the chiral densities,
and the right hand side of (28) would be zero. 
However since we used the continuum determinant no doublers can be expected
and we conclude that the formation of the chiral condensate    
at $e > 0$ is a strong indication that there is only one flavor of 
fermions which has the chiral properties of the continuum 
already for finite correlation length. Also the continuum limit discussed 
in the last section supports this result.

\section{Osterwalder-Schrader positivity}
Osterwalder and Schrader (1973,1975) developed a mathematical procedure
that
allows the reconstruction of the Hamiltonian and the physical Hilbert space 
from a continuum field theory defined in Euclidean space. For the Wilson
formulation of lattice gauge theorie this condition aws shown to hold
Osterwalder and Seiler (1978). For the hybrid approach a new proof has
to be given. 

The main condition in question is the Osterwalder-Schrader 
(OS) positivity. It requires
(see e.g. Glimm \& Jaffe 1987 for an introduction)
\begin{equation}
\langle P_+ \; \Theta P_+\rangle \; \geq \; 0  \; \; \; \; , \; 
\; \; \; \forall P_+ \; ,
\end{equation}
where $P_+$ denotes some operator with support only 
for positive time argument. 
$\Theta$ is the antilinear time reflection 
operator which maps $P_+$ to some operator with support only for negative 
time argument. 

A sufficient condition for OS-positivity is to show that the 
action can be decomposed to (see e.g. Fr\"ohlich et al. 1978)
\begin{equation}
- S \; \; = \; \; - S^+ \; - \; \Theta S^+ \ + \; \int d\mu(\lambda)
C^+(\lambda) \Theta C^+(\lambda) \; ,
\end{equation}
where $S^+$ and $C^+(\lambda)$ contain only field variables with support on 
lattice points with positive time argument and $d\mu(\lambda)$ is some 
positive measure. We will show that the effective action of the Schwinger 
model in the hybrid formulation can be written in the form (30).

We define the time reflection $\Theta$ to be an antilinear operator acting
on the gauge fields as follows
\begin{equation}
\Theta A_1(n_1,n_2) \; := \; A_1(n_1,-n_2+1) \; \; \; , \; \; \;
\Theta A_2(n_1,n_2) \; := \; - A_2(n_1,-n_2) \; .
\end{equation}
This implies for the action of $\Theta$ on the field strength
\begin{equation}
\Theta F_{12}(n_1,n_2) \; = \; -F_{12}(n_1,-n_2) \; .
\end{equation}
It is well known how to decompose the gauge field part of the action into
the form (30) (see e.g. Osterwalder \& Seiler (1978)). 
Thus there is only the part $S_F$ from the fermion action 
given by Equation (10) left to decompose. We work in momentum space and 
rewrite the Fourier transform as
\begin{equation}
\widehat{F_{12}}(p) \; = \; \sum_{n \in Z\hspace{-1.35mm}Z^2} 
e^{-ipn} F_{12}(n) \; = \;
\widehat{F_{12}^+}(p_1,p_2) + \widehat{F_{12}^0}(p_1) -
\Theta \widehat{F_{12}^+}(-p_1,p_2) \; ,
\end{equation}
where (32) and the antilinearity of $\Theta$ were used. We defined
\begin{equation}
\widehat{F_{12}^+}(p_1,p_2) := \!\!\sum_{n_1 \in Z\hspace{-1.35mm}Z^2} 
\!e^{-ip_1 n_1}\!
\sum_{n_2 > 0}\!e^{-ip_2n_2} F_{12}(n_1,n_2) \; , \; \; 
\widehat{F_{12}^0}(p_1) := \!\!\sum_{n_1 \in Z\hspace{-1.35mm}Z^2}\! 
e^{-ip_1 n_1}
F_{12}(n_1,0).
\end{equation}
Making use of the antilinearity of $\Theta$ and (32), one easily verifies 
\begin{equation}
\Theta \widehat{F_{12}^0}(p_1) \; = \; - \widehat{F_{12}^0}(-p_1) \; .
\end{equation}
Inserting the decomposition (33) into (10) one finds
\begin{equation}
-S_F \; = \; -S_F^+ \; - \; \Theta S_F^+ \; - \; S_F^{mix} \; ,
\end{equation}
where 
\begin{equation}
S_F^+ \; = \; -\frac{e^2}{2\pi} \int_{-\pi}^\pi \frac{d^2p}{(2\pi)^2}
\widehat{F_{12}^+}(-p_1, -p_2)  \; \sigma_\mu(p) \; 
\Big[\widehat{F_{12}^+}(p_1, p_2) + 2 \widehat{F_{12}^0}(p_1) \Big]  \; ,
\end{equation}
and
\[
-S_F^{mix} \; := \; \frac{e^2}{2\pi} \int_{-\pi}^\pi \frac{d^2p}{(2\pi)^2}
\widehat{F_{12}^0}(p_1)  \; \sigma_\mu(p) \; \Theta \widehat{F_{12}^0}(p_1)
\]
\begin{equation}
+ \; \frac{e^2}{\pi} \int_{-\pi}^\pi \frac{d^2p}{(2\pi)^2}
\widehat{F_{12}^+}(p_1,p_2) \;  \sigma_\mu(p) \; 
\Theta \widehat{F_{12}^+}(p_1,-p_2)
\; .
\end{equation}
In Equation (36) we used the symmetries of $\sigma(p)$ 
under the interchanges $p_1 \rightarrow -p_1$ and $p_2 \rightarrow -p_2$ 
(compare (39) below). We furthermore introduced an infrared cutoff $\mu$
in order to deal 
in the following with convergent integrals only. In particular we replaced 
$\sigma(p)$ by $\sigma_\mu(p)$ defined as 
\begin{equation}
\sigma(p)_\mu \; := \;
\sum_{{k} \in Z\hspace{-1.35mm}Z^2}
\frac{1}{({p} + 2\pi {k}\,)^2 + \mu^2} \;
\frac{2-2\cos( p_1)}{(p_1 +  2 \pi k_1)^2} \;
\frac{2-2\cos( p_2)}{(p_2 +  2 \pi k_2)^2} \; .
\end{equation}
The limit $\mu \rightarrow 0$ which removes the cutoff will be discussed 
in the end. In order to establish the decomposition (30) of the action, 
there is only left to show that $-S_F^{mix}$ can be written 
in the form of the last term in (30), i.e. as an integral over 
$C^+(\lambda) \Theta C^+(\lambda)$, with some positive measure for $\lambda$. 
Since $\sigma(p)_\mu$ is positive, the first term of $S_F^{mix}$ has already 
this form. The second term can be written as 
\[
\frac{e^2}{\pi} \int_{-\infty}^\infty \frac{d^2p}{(2\pi)^2}
\widehat{F_{12}^+}(p_1,p_2) \frac{2-2\cos(p_1)}{p_1^2}
\frac{2-2\cos(p_2)}{p_2^2} \frac{1}{p^2 + \mu^2} \Theta \widehat{F_{12}^+}
(p_1,-p_2) 
\]
\begin{equation}
= \; \sum_{n_1,m_1 \in Z\hspace{-1.35mm}Z^2} 
\sum_{n_2,m_2 > 0} F_{12}(n_1,n_2) \Theta 
F_{12}(m_1,m_2) M_\mu(n_1-m_1,n_2+m_2) \; .
\end{equation}
We exchanged the sum in $\sigma(p)$ and the integral over
the first Brillouin zone and rewrote sum and integral to an integral
over $\mbox{I\hspace{-0.7mm}R}^2$. Furthermore the
definition (34) of $\widehat{F_{12}^+}(p_1,p_2)$ was inserted and we defined
\[
M_\mu(n_1-m_1,n_2+m_2) :=
\frac{e^2}{\pi} \!\int_{-\infty}^\infty \!\frac{d^2p}{(2\pi)^2}
\frac{2\!-\!2\cos(p_1)}{p_1^2} \frac{2\!-\!2\cos(p_2)}{p_2^2} 
\frac{e^{ip_1(n_1-m_1)} e^{ip_2(n_2+m_2)}}{p^2 + \mu^2}
\]
\begin{equation}
= \; e^2 \int_{-\infty}^\infty \frac{dp_1}{(2\pi)^2}\;
\frac{2-2\cos(p_1)}{p_1^2} \; \frac{2\cosh(p_1) - 2}{p_1^2}\;
\frac{e^{ip_1(n_1-m_1)} e^{-(n_2+m_2)\sqrt{p_1^2 + \mu^2}}}
{\sqrt{p_1^2 + \mu^2}} \; .
\end{equation}
In the second step we solved the $p_2$-integration using a contour integral.
Note that $n_2 + m_2 > 1$ (which can be seen to hold from (40))
is necessary to close the contour in the upper half of the complex 
$p_2$-plane. Putting things together we end up with the following 
expression for the second term in $S_F^{mix}$
\begin{equation}
e^2 \int_{-\infty}^\infty \frac{dp_1}{(2\pi)^2} \;
\frac{2-2\cos(p_1)}{p_1^2}\; \frac{2\cosh(p_1) - 2}{p_1^2} \;
\frac{1}{\sqrt{p_1^2 + \mu^2}} f_\mu (p_1) \Theta f_\mu (p_1) \; ,
\end{equation}
where we defined 
\begin{equation}
f_\mu(p_1) \; := \; \sum_{n_1 \in Z\hspace{-1.35mm}Z} 
\sum_{n_2 > 0}
e^{ip_1n_1} e^{-n_2\sqrt{p_1^2 + \mu^2}} F_{12}(n_1,n_2)\; .
\end{equation}
Expression (42) establishes the decomposition of the action into the 
form (30) for the model with the cutoff. Following Fr\"ohlich 
et al. (1978) this 
implies that (29) holds for the model with the cutoff. Since the covariance 
operator of the cutoff model depends continuously on $\mu$ (replace 
$\sigma(p)$ in (13) by $\sigma_\mu(p)$), so do correlation functions of the 
type (29). Thus we conclude that (29) also holds for the $\mu = 0$ model we 
are interested in, and that the physical Hilbert space can be reconstructed.

\section{Remarks on the chiral Schwinger model}

It has been demonstrated that the vectorlike 
Schwinger model with interpolated gauge
fields in combination with a continuum fermion determinant gives an 
interesting effective lattice gauge theory. It was shown that the model 
has a critical point where the continuum limit can be taken, that the chiral
symmetry is kept intact and that OS-positivity holds. 

The next logical step is to test the hybrid approach in a chiral 
model. For chiral lattice fermions there is the celebrated 
theorem by Nielsen and Ninomya (1981a,b) stating that to each left handed 
fermion the lattice discretization produces a right handed partner, thus
making it impossible to formulate chiral gauge theories on the lattice.   
The hybrid approach offers a chance to overcome this problem. 

For the chiral Schwinger model the fermion determinant in the continuum
can be computed analytically (Jackiw \& Rajaraman 1985). 
It gives rise to an effective action which is not gauge invariant
and complex. Furthermore it contains in addition to the gauge coupling $e$,
a second parameter $a$, parametrizing an arbitrariness in the regularization 
of the short distance singularity showing up in the fermion determinant. 
Inserting 
the interpolated gauge field (4) into the continuum determinant one obtains 
an effective lattice gauge theory depending on two parameters $a$ and $e$. 
It can be shown (Gattringer 1996), that this effective model has a 
critical line in the $e,a$-plane, where the continuum limit can be 
constructed, and that the chiral properties of the continuum model are 
kept intact. Finally, using methods similar to the proof for the vectorlike 
model, OS-positivity can be established as well.
\vskip3mm
\noindent
The author thanks Erhard Seiler for many fruitful discussions.
\vskip2mm
\noindent
{\bf References :}
\vskip1mm
\noindent
L.V. Belvedere, J.A. Swieca, K.D. Rothe, B. Schroer, {\sl Nucl. Phys.}
{\bf B153} (1979) 112; \\
R. Flume, D. Wyler, {\sl Phys. Lett.}{\bf 108 B} (1982) 317; \\
J. Fr\"ohlich, R. Israel, E.H. Lieb, B. Simon, {\sl Commun. Math. Phys.}
{\bf 62} (1978) 1; \\
C. Gattringer, {\sl hep-lat/9511019, to appear in Phys. Rev.} {\bf D}
(1995); \\
C. Gattringer, {\sl work in preparation} (1996); \\
C. Gattringer, E. Seiler, {\sl Ann. Phys.}{\bf 233} (1994) 97; \\ 
J. Glimm, A. Jaffe, {\sl Quantum Physics}, Springer, New York 1987; \\
M. G\"ockeler, G. Schierholz, {\sl Nucl. Phys. (Proc. Suppl.)} {\bf
29B,C} (1992) 114; \\
M. G\"ockeler, G. Schierholz, {\sl Nucl. Phys. (Proc. Suppl.)} {\bf 30B} 
(1993) 609; \\
M. G\"ockeler, A.S. Kronfeld, G. Schierholz, U.-J. Wiese, 
{\sl Nucl. Phys.} { \bf B404} (1993) 839; \\
P. Hernandez, R. Sundrum, {\sl Nucl. Phys.} {\bf B455}( 1995a) 287; \\
P. Hernandez, R. Sundrum, {\sl hep-ph/9510328} (1995b); \\
P. Hernandez, R. Sundrum, {\sl hep-lat/9602017} (1996); \\
G. 't Hooft, {\sl Phys. Lett.} {\bf 349 B} (1995) 491; \\
S.D.H. Hsu, {\sl hep-th/9503058} (1995); \\
R. Jackiw, R. Rajaraman, {\sl Phys. Rev. Lett.} {\bf 54} (1985) 1219; \\
M. L\"uscher, {\sl Commun. Math. Phys.} {\bf 85} (1982) 39; \\
I. Montvay, {\sl hep-lat/9505015} (1995); \\
H.B. Nielsen, M. Ninomya, {\sl Nucl. Phys.} {\bf B185} (1981a) 20; \\
H.B. Nielsen, M. Ninomya, {\sl Nucl. Phys.} {\bf B193} (1981b) 173; \\
K. Osterwalder, R. Schrader, {\sl Commun. Math. Phys.} {\bf 31} (1973) 83; \\
K. Osterwalder, R. Schrader, {\sl Commun. Math. Phys.} {\bf 42} (1975) 281; \\
K. Osterwalder, E. Seiler, {\sl Ann. Phys.} {\bf 110} (1978) 440; \\
J. Schwinger, {\sl Phys. Rev.} {\bf 128} (1962) 2425; \\
Y. Shamir, {\sl hep-lat/9509023} (1995) ; \\
R.L. Stuller, {\sl Hybrid Quantization} (unpublished notes); \\
\end{document}